\documentclass[aps,pre,reprint,superscriptaddress]{revtex4-1}
\usepackage{amsmath,amssymb}
\usepackage{graphicx}

\usepackage[colorlinks,linkcolor=blue,citecolor=blue,urlcolor=blue]{hyperref}

\begin{document}

\title{On Target Pattern Formation in the CHNS system}
\author{Qinghao Yan}
\email{qinghaoyan@outlook.com}
\affiliation{Center for Fusion Sciences, Southwestern Institute of Physics, Chengdu, Sichuan 610041, People's Republic of China}
\affiliation{Department of Engineering Physics, Tsinghua University, People's Republic of China}
\author{Patrick H. Diamond}
\affiliation{Center for Fusion Sciences, Southwestern Institute of Physics, Chengdu, Sichuan 610041, People's Republic of China}
\affiliation{CASS and Department of Physics, University of California, San Diego}

\email{diamondph@gmail.com}

\begin{abstract}
		We study the concentration field in a prescribed 2D Cahn-Hilliard 
		Navier-Stokes (CHNS) system. We formulate a description for the target 
		pattern formation and pattern merging processes, and compare this 
		description with simulation results. Shear-augmented diffusion along 
		streamlines causes a separation of time scales, thus 2D CHNS system can 
		be simplified to a 1D system. In this 1D system, target pattern formation is 
		induced by linear instability. The waveform of patterns are described by 
		Jacobi Elliptic Functions. The interface (of pattern) migration or coarsening 
		velocity is determined by the derivative of interface curvature. The 
		anomalous migration of inner pattern can be explained by the singularity 
		at the origin  and therefore the boundary motion in the 
		quasi-one-dimension system. Finally we derive a simple criterion for  when 
		CHNS system becomes dynamic by following similar cases in MHD.
\end{abstract}
\maketitle
\section{Introduction}
In a miscible binary liquid mixture analogy to water and alcohol, with temperature below a critical value,  the phases of the fluid could evolve to immiscibility like water and oil. This process is named as spinodal decomposition. The 2D CHNS system is a standard model for studying spinodal decomposition, which combines the Cahn-Hilliard equation and Navier-Stokes equation as below. 
\begin{gather}
\dfrac{\partial\psi}{\partial t}+\mathbf{u}\cdot\nabla\psi=D\nabla^2[-\psi+\psi^3-\xi^2\nabla^2\psi]\label{originalCH}\\
\partial_t\omega+\mathbf{u}\cdot\nabla\omega=\dfrac{\xi^2}{\rho}\mathbf{B}_{\psi}\cdot\nabla\nabla^2\psi+\nu\nabla^2\omega\label{vorticity}\\
\mathbf{u}=\hat{z}\times\nabla\phi,\quad \omega=\nabla^2\phi\\
\mathbf{B}_{\psi}=\hat{z}\times\nabla\psi,\quad j_{\psi}=\xi^2\nabla^2\psi
\end{gather}
In the first stage of spinodal decomposition, one of the components will form small droplets. This stage is called phase separation and well described by highly nonlinear model\cite{Mauri96}. An analytical solution like $\tanh(\sqrt{(x-x_0)^2+(y-y_0)^2}/(\sqrt{2}\xi))$ could be used to describe the interface between two different components. Obviously, the parameter $\xi$ gives the width of the interface. Then during the second stage of spinodal decomposition, the phase coarsening, small droplets will merge to big droplets. The steady size of bubble is determined by the balance between elastic energy and turbulent kinetic energy.  And this size can be described by the Hinze scale $L_H=(\rho/\sigma)^{-3/5}\varepsilon^{-2/5}$, where $\rho$ is the density, $\sigma$ is the surface tension and $\varepsilon$ is the dissipation rate per unit mass\cite{Hinze55, Perlekar12, Perlekar14}. Under some conditions, C-H equation converges to Mullins-Sekerka problem\cite{Cahn96, Pego261} or Hele-Shaw model\cite{Alikakos94}. In both model, the motion of interfaces are related to the derivative of mean curvature of interfaces. Because of the similarities of control equations between CHNS and 2D MHD system, some studies put them together for comparison\cite{Ruiz81, Fan16, Perlekar14}. Different simulation methods are applied for C-H flow coarsening too\cite{Yue04, Zhu99}.\par
\begin{figure}
\centering
\includegraphics[width=0.75\linewidth]{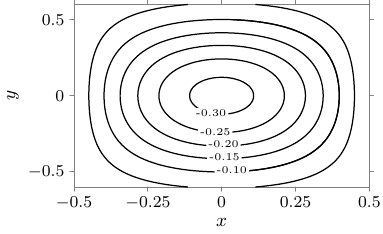}
\caption{The target-like velocity potential $\phi/\phi_0$, and thus target-like velocity field.}
\label{F:phi}
\end{figure}

For passive flow, which means the flow field isn't influenced by the velocity field, we can ignore the second equation (\ref{vorticity}) and just set a background velocity. Depending on what we focus on, the velocity field can be different. Examples in MHD are Ref.\cite{Moffatt83, Weiss66} for Flux Expulsion. If we want to study the shearing effect of velocity, we can set $\mathbf{u}=u(y)\hat{x}$ and  $\psi(x,y,t)=A(y,t)\cos k(x-ut)$ for the equation (\ref{originalCH}). Keep the highest power of time $t$ only. The rapid decay process will give a time scaling, $t_{\text{FE}}=\text{Rm}^{1/3}t_0, \text{Rm}\gg1$ for MHD,  where $t_0$ is the eddy turn over time, $\text{Rm}$ is the magnetic Reynolds number.  Meanwhile, for CHNS the time scale is $t_{\text{CHNS}}=\text{Rm}_{\text{e}}^{1/5}t_0$, where $\text{Rm}_{\text{e}}$ is the effective Reynolds number defined in CHNS analogy to $\text{Rm}$. Such time scale indicates that the fastest evolution process of $A(y,t)$ is
\begin{equation}\label{E:A0}
	A(y,t)=A_0\exp\bigg(\dfrac{-D\xi^2(ku_{y})^4t^5}{5}\bigg)\xrightarrow{t\rightarrow\infty} 0
\end{equation}
It shows the \emph{shear augmented diffusion}\cite{Rhines83} effect of $u_y$, which derive the concentration field to 0 rapidly. Also note that if $u_y\equiv\text{d}u/\text{d}y=0$, then $A(y,0)=A_0$. Another interesting example is setting a differential rotating flow field, like Goldreich and Lynden-Bell \cite{Goldreich65} did to study galaxy arm formation, or Weiss \cite{Weiss66} did to study MHD flux expulsion. The velocity potential of such eddy flow are like below. The shape of this eddy is quite like a rounded rectangle or \emph{target}, but fundamentally closed streamlines, see FIG. \ref{F:phi}. 
\begin{equation}
\phi=-\dfrac{\phi_0}{\pi}(1-4y^2)^4\cos(\pi x)
\end{equation}

\begin{figure}
	\centering	\includegraphics[width=\linewidth]{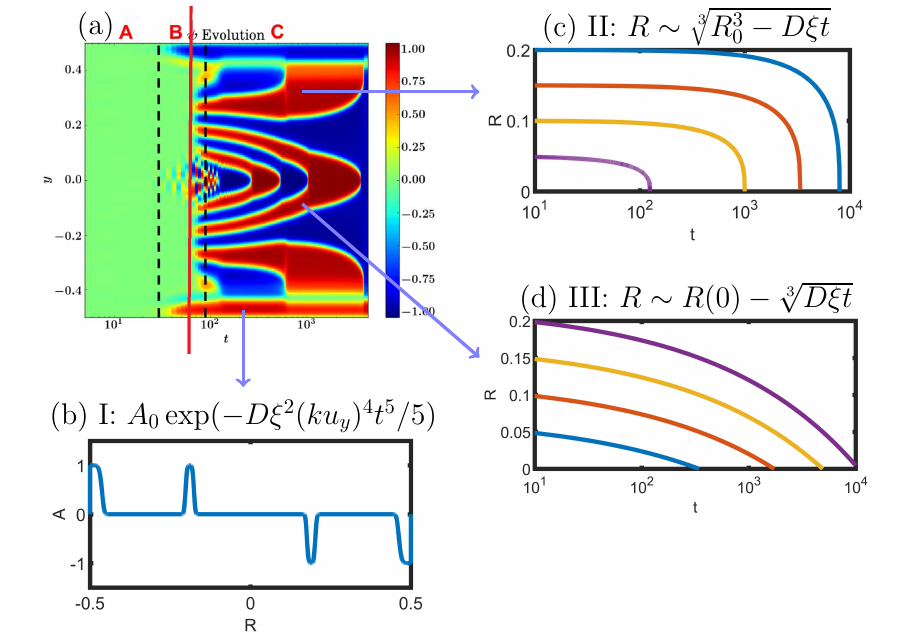}
	\caption{Formation and evolution of target pattern in CHNS system. (a) Reprint from Fan \cite{Fan17}, the simulation result of KS equation in a double-periodical boundary cell, target patterns form near $t=10^2$ and merge.  (b) Rapid decay described by (\ref{E:A0}). (c) Trajectories of interfaces in Region-II, which is determined by the derivative of interface mean curvature (\ref{E:VII}). (d) Trajectories of interface in Region-III, which has the background velocity comes from the singularity of origin (\ref{E:RIII}). }
	\label{F:evolution}
\end{figure}\par
The most interesting phenomenon in CHNS system with such prescribed velocity field is the formation of the \emph{target pattern}\cite{Fan17}. Here is an example  in Ref. \cite{Fan17} Fig. \ref{F:evolution}(a). During the early stage of evolution, when look through the red vertical line, the concentration field have non-zero values near the boundaries (where $u_y\approx0$, let's call it region I), while zero values in other domains (see FIG. \ref{F:evolution}(b)). Then about $ t\gtrsim 10^2$, target patterns appear.
 In general CHNS simulations, during the spinodal decomposition, the system will contains a lot of droplets. So target pattern is actually  a regulated state of these droplets, and will merge to larger pattern like small droplets merging to big droplets.  The  back reaction of amplified field to eddy will lead to the disruption of the eddy and therefore effect the turbulence. In MHD, this reflects in the modification of transport coefficient. Therefore, studying the non-passive condition of CHNS system might enlighten the nature of spinodal decomposition in turbulence.\par
In this study, we use curvilinear coordinates to simplify the 2D CHNS system to 1D CH system, then apply linear instability analysis to study the early stage of target pattern formation in Sec.\ref{S:2}. In Sec.\ref{S:3} we describe the process of interfaces motion and pattern merging. In Sec.\ref{S:4} we derive a simple criterion for system becoming dynamic. Conclusion and discussion are displayed in Sec.\ref{S:5}.\par
\section{Target Formation\label{S:2}}
\quad As mentioned before, the flow $\mathbf{u}$ is prescribed, we only need to consider the the C-H equation. Because the fastest process is the averaging of $\psi$ along the streamlines (caused by shear augmented diffusion)  \cite{Rhines83}, after \emph{averaging} there should be no variation of $\psi$ along the streamlines: $\nabla\psi=\nabla_{\phi}\psi\mathbf{g}_{\phi}$. Therefore, it's reasonable to represent the C-H equation in curvilinear coordinates $(\phi, s)$ for simplicity.  What we expect is a simplified 1D C-H equation solved by a periodic solution.

\subsection{One Dimensional Analysis\label{S:2b}}
\begin{figure}
\centering
\includegraphics[width=0.9\linewidth]{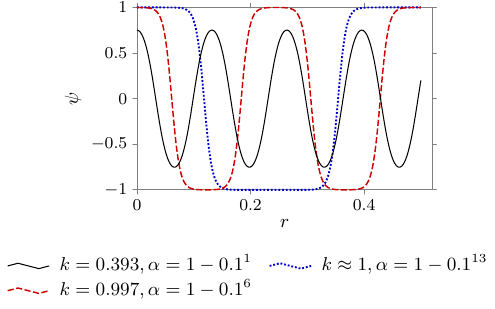}
\caption{There exits a set of solutions $\psi(r,k)=\beta \text{sn}\bigg(\dfrac{\alpha}{\beta}\dfrac{1}{\sqrt{2}\xi}(r+\dfrac{\lambda}{4}),k\bigg)$ with different $\alpha$ and satisfy the equation (\ref{E:9}).}
\label{F:3}
\end{figure}

The transformed C-H equation in curvilinear coordinates $(\phi, s,z)$ is (see appendix \ref{Appendix})
\begin{equation}\label{E:5}
\dfrac{\partial\psi}{\partial t}=D\Delta(-\psi+\psi^3-\xi^2g^{11}\psi_{,\phi\phi}-\xi^2\kappa\psi_{,\phi})
\end{equation}
The bulk of expressions in the bracket is the chemical potential $\mu(\psi)=\mu_0(\psi)+\mu_1(\psi)+\text{etc.}$. The last term is of the order $\text{O}(\xi)$, while others are $\text{O}(1)$.  
For the lowest order, we aim to find a quasi-static state, absorb the coefficient $g^{11}$, and define an effective radius variable $r$. Then there results one dimensional ODE from the chemical potential as below, whose solutions are the C-H equation's lowest-order-equation solutions. 
\begin{equation}\label{E:9}
		-\psi+\psi^3-\xi^2\dfrac{\partial^2\psi}{\partial r^2}=0
\end{equation}
A particular solution of this ODE is $\psi=\tanh[r/\sqrt{2\xi^2}]$, which gives the typical structure of the interface \cite{Yavuz08}. And this corresponds to the absolute minimum of the free energy \cite{Amy08}. However it's not periodic. The periodic solution should be an even function $\psi_{r}(0)=0$. Let the boundary condition be $$\psi_{r}(\pm0.5)=\dfrac{0.5\alpha }{\sqrt{2}\xi},\alpha\le1$$We suppose the solution has the Jacobi Elliptic Function form $A\text{sn}(Br,k)$ \cite{Mauri96}. Put it into the equation (\ref{E:9}), we get the solution
\begin{equation}\label{E:periodical}
	\psi=\beta \text{sn}\bigg(\dfrac{\alpha}{\beta}\dfrac{1}{\sqrt{2}\xi}(r+\dfrac{\lambda}{4}),k\bigg)
\end{equation}
Here, $\beta=\sqrt{1-\sqrt{1-\alpha^2}}$, $k^2=(2\beta^2/\alpha^2)-1$. The shift of $\lambda/4$ in $r$ is to satisfy the requirement of $\psi$ being an even function. Sn is a periodic function, and the period of $\psi$ is defined by
\begin{equation}
	\lambda=\dfrac{4 \sqrt{2}\beta\xi}{\alpha}\int_0^{\pi/2}\dfrac{\text{d}\varphi}{\sqrt{1-k^2\sin\varphi}}
\end{equation}
When $k\rightarrow1$, $\lambda\rightarrow\infty$, sn tends to a $\tanh$. When $k=0$, sn has a minimum period $\lambda_{\min}=(2\sqrt{2}\pi\beta\xi)/\alpha$. So there is a range of period, $\lambda\in[\lambda_{min},\infty)$, see FIG. \ref{F:3} . There must be a $\lambda_s$ corresponding to the quasi-steady state we want.
\par
A standard method can be used to get the initial period of target pattern, i.e. linear instability analysis. The basic idea is the $\lambda$ which makes the instability grow fastest is the one to form the target pattern\cite{Mauri96,Amy08}.\par
The linearized C-H equation is 
\begin{equation}
	\psi_t=D[(-1+3\bar{\psi}^2)\nabla^2\tilde{\psi}-\xi^2\nabla^4\tilde{\psi})]
\end{equation}
Here $\bar{\psi}$ is the \emph{homogeneous steady state}, $\tilde{\psi}$ is the perturbation value.  Assuming the perturbation is 
$\tilde{\psi}=\delta\psi_0\cos(k_y y)e^{\sigma t}$. Then the k-mode growth rate is
\begin{equation}\label{E:growthrate}
	\sigma=D[(1-3\bar{\psi}^2)k_y^2-\xi^2k_y^4]
\end{equation}
It's obvious that $\sigma$ depends on the homogeneous steady state. In the previous discussion (FIG. \ref{F:evolution}(b)), we showed that the fastest process in this system is shear-augmented diffusion (\ref{E:A0}). This diffusion causes the value of $\psi$ rapidly to decay to zero for those areas that $u_y\neq 0$. But for areas with $u_y=0$ and near the up and down boundary, $\psi=\pm1$. Since target pattern formation shows up after the rapid decay, these homogeneous states will determine the linear instability growth rate. In the first case, $\bar{\psi}=\pm1$, and $\sigma<0$, the perturbed value with any $k$ won't grow, and system will keep the initial homogeneous steady state.\par
In the second case, $\bar{\psi}=0$, then 
\begin{equation}
	\sigma=Dk_y^2(1-\xi^2k_y^2)
\end{equation}
And when $k_y=k_{ys}=1/(\sqrt{2}\xi)$, we get the maximum growth rate $\sigma$, and the corresponding $\lambda_s=2\sqrt{2}\pi\xi$. If $ \xi=10^{-2}$, then $\lambda_s\approx0.089$.  Returning to the length scale before normalization, the period of the original pattern is $\lambda_s\approx0.089L_0$, where $L_0$ is the characteristic length of the eddy.\par
In the simulation of Fan (Ref.\cite{Fan17}), also see FIG. \ref{F:evolution}, the Region-I patterns form near up and down boundaries  at $t\sim 20$. And these patterns will keep steady during the whole simulation. When time comes to a bit less than $10^2$, system generate new patterns in the space domain of $y\in[-0.4, 0.4]$. The number of positive patterns is about 9, and number of negative patterns is 8. Basically there are 8 space periods for target pattern. So the wave length of pattern is approximately $\lambda\approx 0.1\sim\lambda_s$. Last but not least, recall the boundary condition we set for the periodic solution of target pattern, $\psi_r|_{\text{boundary}}=\text{constant}$. Since the Region-I patterns are steady than the target patterns in a long time, the out boundary condition of target patterns has some leeway to meet the mathematical consideration. As a conclusion, we can say the results of linear instability analysis fit well with the simulation. \par

\section{Interfaces Motion\label{S:3}}
The next question is how the interfaces move. Since the interfaces and pattern formation are the results of the leading order of chemical potential in (\ref{E:5}), the answer of this problem should be found in the remain part of chemical potential. A matched asymptotic expansion method was applied to C-H system years ago, see Ref. \cite{Pego261,Cahn96}. For such constant-mobility C-H system, the interfaces velocity is described by Mullins-Sekerka problem in both Dirichlet and Neumann type boundary conditions. Because of the result in (\ref{E:A0}), the Region-II patterns have Dirichlet boundary condition. Thus the motion of interface is determined by equations below,
\begin{gather}\label{E:VII}
	V(x,t_1)=D[\hat{n}\cdot\nabla \mu_1]^+_-[\psi]^{-1}\\
	\mu_1=-\kappa(x,t_1)S[\psi]^{-1}, x\in \Gamma\\
	S=\int \psi'(\rho)^2\text{d}\rho\\
	[\psi]=\psi^+-\psi^-
\end{gather}
where $t_1=\xi t$. In this problem, we can use the typical solution $ \psi=\tanh \rho $ to get an intuitive impression about how the interfaces move.  
\begin{align}
	S=\int \psi'(\rho)^2\text{d}\rho\approx\int_{-\infty}^{\infty}\bigg(1-\tanh^2(\rho)\bigg)^2\text{d}\rho\\\notag
	 \lessapprox\int_{-\infty}^{\infty}\bigg(1-\tanh^2(\rho)\bigg)\text{d}\rho=\tanh(\rho)\bigg\vert^{\infty}_{-\infty}=[\psi]
\end{align}
which leads to $\mu_1=-\kappa$.
Let's see the simplest case. Set $\kappa=1/R$, obtain the $V$, and calculate the interfaces trajectories (see FIG. \ref{F:evolution}(c)). 
The trajectories of interfaces are similar to Region-II patterns in simulation.
The Region-II and Region-III patterns move a bit different, but the time scales are the same. To explain the Region-III patterns' motion, we have to consider more about the boundary condition. We should notice that when the near origin patten interface vanishes, the outer interfaces will suddenly drop to the origin, while the other patterns don't. This indicate that the motion of interfaces are related to the inner interfaces, and to the inner boundary. We assume the boundary of near origin interface is actually moving, we should account for the boundary (origin) effect. And the motion law could be,
\begin{equation}
	V(x,t)\sim V_{\text{O}}
\end{equation}
Here O is origin for short, $V_{\text{O}}$ is the background speed causing by the singularity of origin. We suppose that $V_{\text{O}}$ follows the same motion law like interfaces (\ref{E:VII}), and just set $ R_\text{O} =0$,
\begin{equation}\label{E:VIII}
	V_{\text{O}}(t)\sim -\dfrac{D\xi}{\sqrt[3]{(R_{\text{O}}^3-D\xi t)^2}}\sim- \dfrac{D\xi}{\sqrt[3]{(D\xi t)^2}}
\end{equation}
Then for Region-III patterns, approximately (see FIG.  \ref{F:evolution}(d)),
\begin{equation}\label{E:RIII}
 R_{\text{III}}\sim R_{\text{TR}}^0-\sqrt[3]{D\xi t}
\end{equation}

\section{The Non-passive Condition\label{S:4}}
All the above discussions are based on the condition of \emph{passive flow}. Here we give a simple estimation of when the passive flow condition fails. The main idea is to find out when the dynamical feedback is strong enough to violate the kinematic approach.\par
In 2D MHD, there exits freezing-in-law due to the induction equation of magnetic field. Since the CHNS system is similar to 2D MHD, similar phenomenon should exist. The effect of $\mathbf{B}_{\psi}\cdot\nabla\mathbf{U}$ in an eddy causes field amplification. When this field amplification is comparable to the hyper-diffusion term, amplification will stop\cite{Mak17}.
\begin{equation}
	|\mathbf{B}_{\psi}\cdot\nabla\mathbf{U}|\sim|D\xi^2\nabla^4\mathbf{B}_{\psi}|
\end{equation}
The initial value of the effective magnetic field $B_{\psi}$ is $B_0$, stretched value is $b$. Assuming after the flux expulsion process, the field are almost expelled in a narrow width $l$. The characteristic velocity and length of eddy are $U_v$ and $L_v$. So the balance of hyper-diffusion and stretching can be written as
\begin{equation}
	B_0\dfrac{U_v}{L_v}\sim D\xi^2\dfrac{b}{l^4}
\end{equation}
Plus the flux conservation relation $B_0L_v=bl$, we obtain,
\begin{equation}\label{E:scales in NP}
	\begin{cases}
		b=B_0\bigg(\dfrac{U_vL_v^3}{D\xi^2}\bigg)^{\frac{1}{5}}=B_0\text{Rm}_{\text{e}}^{\frac{1}{5}}\\
		l=L_v^{\frac{2}{5}}\bigg(\dfrac{D\xi^2}{U_v}\bigg)^{\frac{1}{5}}=L_v\text{Rm}_{\text{e}}^{-\frac{1}{5}}
	\end{cases}
\end{equation}
The $\text{Rm}_{\text{e}}$ is the effective magnetic Reynolds number, defined using hyper-diffusion term.\par
Field amplification increases the effective magnetic tension, which we can estimate using the decomposition in Ref. \cite{Mak17}. $\mathbf{B}_{\psi}$ is mostly in $\mathbf{g}^2$ direction $\mathbf{B}_{\psi}\approx B_s\mathbf{g}^2$. We assume the coordinates are orthogonal, then $\mathbf{g}^2$ is parallel to $\mathbf{g}_2$.
\begin{align}\label{E:BpsiTension}
	|\mathbf{B}_{\psi}\cdot\nabla\mathbf{B}_{\psi}|&\sim |B_s B_{s;j}\mathbf{g}^{2}\cdot(\mathbf{g}^j\mathbf{g}^2)|\\\notag
	&\sim\bigg|-\dfrac{|\mathbf{B_{\psi}}|^2}{L_v}\mathbf{e}_{\phi}+\dfrac{\text{d}}{\text{d}s}(\dfrac{|\mathbf{B}_{\psi}|^2}{2})\mathbf{e}_{s}\bigg|\\\notag
	&\sim \dfrac{b^2}{L_v}
\end{align}
This is the maximum value of effective magnetic tension. For vorticity equation (\ref{vorticity}), if this increased tension can compete with the advection of vorticity, we can say the eddy is disrupted. 
\begin{equation}\label{E:disrupted}
	|\mathbf{U}\cdot\nabla\mathbf{\omega}|\sim\bigg|\dfrac{\xi^2}{\rho}\nabla\times(\mathbf{B}_{\psi}\cdot\nabla\mathbf{B}_{\psi})\bigg|
\end{equation}
For a vector in curvilinear coordinates, $\nabla\times \vec{v}=\varepsilon^{ijk}\nabla_i v_j \mathbf{g}_k$, then, 
\begin{equation}\label{E:curTension}
 |\nabla\times(\mathbf{B}_{\psi}\cdot\nabla\mathbf{B}_{\psi})|=\sqrt{g}\bigg| \dfrac{\partial}{\partial\phi}\bigg(\dfrac{\text{d}}{\text{d}s}\dfrac{|\mathbf{B}_{\psi}|^2}{2}\bigg)+\dfrac{\text{d}}{\text{d}s}\dfrac{|\mathbf{B}_{\psi}|^2}{L_v}\bigg|\sim \dfrac{b^2}{L_v l}
\end{equation}
Since the stretched field is locally, the transverse derivative $\sqrt{g}\partial/\partial\phi$ will take $l$ as the characteristic length. In the left hand side of (\ref{E:disrupted}), gradient of vorticity is estimated as $U_v/L_v^2$. Together with equation (\ref{E:scales in NP}) and  (\ref{E:curTension}), obtains
\begin{equation}\label{E:Rmh}
	U_v\dfrac{U_v}{L_v^2}\sim \dfrac{\xi^2}{\rho}\dfrac{b^2}{lL_v}\notag\Longrightarrow\dfrac{\xi^2B^2}{\rho U_v^2}\text{Rm}_{\text{e}}^{3/5}\sim 1\notag \Longrightarrow
	M^2\text{Rm}_{\text{e}}^{3/5}\sim 1
\end{equation}
Here $M$ is the ratio of Alfv\'{e}n speed and eddy characteristic speed, or the measuring of the filed strength. Compare with the result in MHD,
\begin{equation}
		M^2\text{Rm}\sim1
\end{equation}
They have the same structure. We can rewrite (\ref{E:Rmh}) using the P\'{e}clet number $\text{Pe}=L_vU_v/D$ as 
\begin{equation}
	M^2\text{Pe}\bigg(\dfrac{L_v^2 D}{U_v\xi^3}\bigg)^{2/5}\sim1
\end{equation}

\section{Conclusion\label{S:5}}
We study the CHNS system with prescribed eddy-like velocity field, the formation process of target pattern is studied. The separation of time scale is the key point to simplify the system. The time scale of shear-augmented diffusion is longer than the eddy turn over time, and causes the homogeneity along each streamlines. Under the condition of passive flow, we simplify the 2D CHNS system to a 1D CH system. As a mathematical problem, one of the periodic solutions of 1D Cahn-Hilliard equation is the Jacobi Elliptic Function. From linear instability analysis, we know different homogeneous states will evolve differently. Near-boundary pattern will stay steady. Inner regions will evolve the target patterns, the number of initial patterns is given by the most unstable mode. Then in inner regions, the interfaces of target pattern will diffuse towards origin slowly, and merge at origin. So the number of patterns will decrease. The motion speed is determined by the derivative of curvature of interface. Because of the singularity of origin, the motion speed is accompanied by a background velocity. Finally, we discuss when the system becomes dynamical, following the similar procedures in MHD. The non-passive condition is $M^2\text{Rm}_{\text{e}}^{3/5}\sim 1$. This implicates that the transport coefficient in spinodal turbulence might relate to this dynamical relation in the same way as in MHD.

\begin{acknowledgments}
	We thank Zhibin Guo and Xiang Fan for helpful discussion. This work is supported by the National Key Research and Development Program of China under Grant No. 2018YFE0309103,  the National Natural Science Foundation of China under Grant No. 11875124, U1867222,  11575055 and  11705052.  The work is also supported by the U.S. Department of Energy, Office of Science, Office of Fusion Energy Sciences under Award Number DE-FG02-04ER54738.
\end{acknowledgments}

\appendix
\section{Transformation of Coordinates\label{Appendix}}
\begin{figure}
	\centering
	\includegraphics[width=0.9\linewidth]{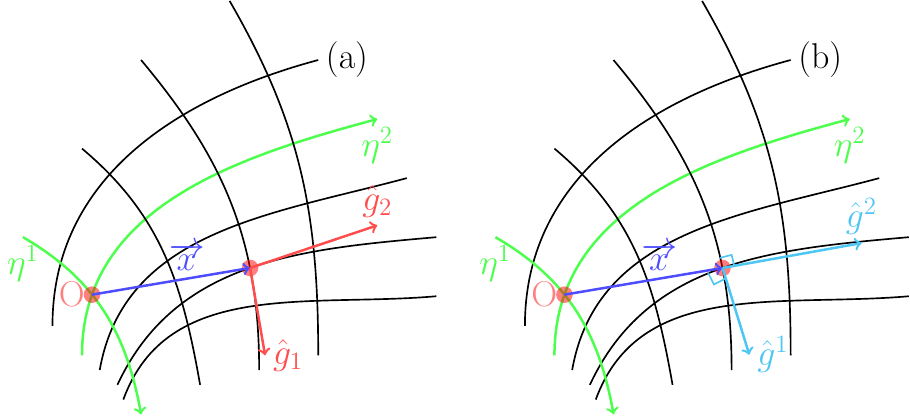}
	\caption{Dual Bases. (a) Covariant bases, (b) Contravatiant bases.}
	\label{F:dual}
\end{figure}

First, introduce some basic informations about dual bases, see FIG. \ref{F:dual}. For a point with curvilinear coordinates $(\eta^1,\eta^2,\eta^3)$ identify the position of $\vec{x}$ in space.  The corresponding vectors are $(\mathbf{g}_1,\mathbf{g}_2,\mathbf{g}_3)$. These vectors are called covariant bases, and point to the direction of $\eta^i$ increasing, see FIG. \ref{F:dual}(a). Thus they are tangent to the grid lines. But it's worth to notice $\mathbf{x}\neq\sum \eta^i\mathbf{g}_i$, actually it's $\text{d}\vec{x}=\sum \text{d}\eta^i\mathbf{g}_i\equiv\text{d}\eta^i\mathbf{g}_i$, which use the Einstein summation notation.
\begin{equation}
\mathbf{g}_i\equiv\dfrac{\partial\vec{x}}{\partial\eta^i}
\end{equation}\par
Meanwhile, there exits contravariant basis, which are normal to surfaces of constant $\eta^j$, and point to the direction of $\eta^j$ increasing, see FIG. \ref{F:dual}(b).
\begin{equation}
\mathbf{g}^j\equiv\dfrac{\partial\eta^j}{\partial \vec{x}}
\end{equation}
It's easy to see $\mathbf{g}_i\cdot\mathbf{g}^j=\delta_i^j$.
\par
The derivative of a vector $\mathbf{F}=F^i\mathbf{g}_i$ to $x^j$ is
\begin{align}
\dfrac{\partial\mathbf{F}}{\partial x^j}=\dfrac{\partial F^i}{\partial x^j}\mathbf{g}_i+F^i\dfrac{\partial \mathbf{g}_i}{\partial x^j}=\dfrac{\partial F^i}{\partial x^j}\mathbf{g}_i+F^i\Gamma^k_{ij}\mathbf{g}_k\\\notag
=\bigg(\dfrac{\partial F^i}{\partial x^j}+F^m\Gamma^i_{jm}\bigg)\mathbf{g}_i\equiv F^i_{;j}\mathbf{g}_i
\end{align}

Vector $\mathbf{F}$ could also be written in contravariant bases $\mathbf{F}=F_i\mathbf{g}^i$, and the derivative with respect to $x^j$ is
\begin{equation}
\dfrac{\partial\mathbf{F}}{\partial x^j}=\bigg(\dfrac{\partial F_i}{\partial x^j}-F_m\Gamma^m_{ij}\bigg)\mathbf{g}^i\equiv F_{i;j}\mathbf{g}^i
\end{equation}
Here we can use $F^i_{,j}$ to represent the usual partial differential $\partial F^i/\partial x^j$.  And $\Gamma^k_{ij}\equiv \mathbf{g}^k\cdot (\partial \mathbf{g}_i/\partial \eta^j)$ is the Christoffel symbol of the second kind. 
The Laplace-Beltrami operator is defined as below \cite{Abraham88,Brakke78}, the reason it's called "Laplace-Beltrami" is indicating the curvature might involve in.
\begin{equation}
\Delta F\equiv\nabla^2F=\dfrac{1}{\sqrt{g}}\dfrac{\partial}{\partial x^j}\bigg(g^{ij}\sqrt{g}\dfrac{\partial f}{\partial x^i}\bigg)
\end{equation}
where $g^{ij}$ are the components of metric matrix $G=J^TJ$, $g=|\text{det}(G)|$. \par
Now back to C-H equation, in $(\eta^1,\eta^2)=(\phi, s)$ coordinates. Recall the hypothesis of averaging along streamlines, which leads to simplification, i.e.
\begin{gather}
\nabla \psi=\dfrac{\partial\psi}{\partial \phi}\mathbf{g}^1=\psi_{,\phi}\mathbf{g}^1\\
\mathbf{u}\cdot\nabla \psi=(\hat{z}\times\nabla\phi)\cdot\mathbf{g}^1 \psi_{,\phi}=(\hat{z}\times\mathbf{g}^1)\cdot\mathbf{g}^1|\mathbf{u}|\psi_{,\phi}=0
\end{gather}
Apply the same simplification to Laplace operator, we obtain
\begin{gather}
\Delta \psi=g^{11}\psi_{,\phi\phi}+(g^{1j}\dfrac{\partial\ln\sqrt{g}}{\partial x^j}+\dfrac{\partial g^{1j}}{\partial x^j})\psi_{,\phi}
\end{gather}
Define $$g^{1j}\dfrac{\partial\ln\sqrt{g}}{\partial x^j}+\dfrac{\partial g^{1j}}{\partial x^j}\equiv \kappa$$ then, 
\begin{equation}
\Delta \psi=g^{11}\psi_{,\phi\phi}+\kappa\psi_{,\phi}
\end{equation}
Finally, we obtain the transformed C-H equation in $(\phi, s,z)$ ,
\begin{equation}\label{A:5}
\dfrac{\partial\psi}{\partial t}=D\Delta(-\psi+\psi^3-\xi^2g^{11}\psi_{,\phi\phi}-\xi^2\kappa\psi_{,\phi})
\end{equation}

\bibstyle{apsrev4-1}

\bibliography{cheq}

\end{document}